\documentclass[aps,prb,twocolumn,superscriptaddress,bibliography]{revtex4-2}

\usepackage{graphicx}
\usepackage{dcolumn}
\usepackage{bm}
\usepackage{hyperref}
\usepackage{soul}
\usepackage{graphicx, multirow, xcolor, bm, ulem}
\usepackage{amsfonts,amssymb,amsmath}
\usepackage{graphicx,dcolumn,bm,color,braket,slashed}
\usepackage{times, comment, mathtools, extarrows,textcomp,ulem}

\newcommand{\blue}[1]{{\textcolor{blue}{#1}}}

\newcommand{\BCAO} {BaCo$_2$(AsO$_4$)$_2$}

\begin{document}

\title{Quantum Order by Disorder: A Key to Understanding the Magnetic Phases of \BCAO}

\author{Sangyun Lee}
\thanks{Present Address: Department of Physics and National High Magnetic Field Laboratory, University of Florida, Florida 32611, USA.}
\affiliation{National High Magnetic Field Laboratory, Los Alamos National Laboratory, Los Alamos, New Mexico 87545, USA.}

\author{Shengzhi Zhang}
\affiliation{National High Magnetic Field Laboratory, Los Alamos National Laboratory, Los Alamos, New Mexico 87545, USA.}

\author{S. M. Thomas}
\affiliation{Los Alamos National Laboratory, Los Alamos, NM 87545, USA.}

\author{L. Pressley}
\affiliation{Oak Ridge National Laboratory, Oak Ridge, TN 37830, USA.}

\author{C. A. Bridges}
\affiliation{Oak Ridge National Laboratory, Oak Ridge, TN 37830, USA.}

\author{Eun Sang Choi}
\affiliation{National High Magnetic Field Laboratory, Florida State University, Tallahassee, Florida 32310-3706,  USA.}

\author{Vivien S. Zapf}
\email[Corresponding authors:]{vzapf@lanl.gov}
\affiliation{National High Magnetic Field Laboratory, Los Alamos National Laboratory, Los Alamos, New Mexico 87545, USA.}

\author{Stephen M. Winter}
\email[Corresponding authors:]{winters@wfu.edu}
\affiliation{Department of Physics and Center for Functional Materials, Wake Forest University, Winston-Salem, NC 27109, USA.}

\author{Minseong Lee}
\email[Corresponding authors:]{ml10k@lanl.gov}
\affiliation{National High Magnetic Field Laboratory, Los Alamos National Laboratory, Los Alamos, New Mexico 87545, USA.}

\begin{abstract}
\BCAO{} (BCAO), a honeycomb cobaltate, is considered a promising candidate for materials displaying the Kitaev quantum spin liquid state. This assumption is based on the distinctive characteristics of  Co$^{2+}$ ions (3$d^7$) within an octahedral crystal environment, resulting in spin-orbit-coupled $J_{\rm eff}$~=~1/2 doublet states. However, recent experimental observations and theoretical analyses have raised questions regarding this hypothesis. Despite these uncertainties, reports of continuum excitations reminiscent of spinon excitations have prompted further investigations. In this study, we explore the magnetic phases of BCAO under both in-plane and out-of-plane magnetic fields, employing dc and ac magnetic susceptibilities, capacitance, and torque magnetometry measurement. Our results affirm the existence of multiple field-induced magnetic phases, with strong anisotropy of the phase boundaries between in-plane and out-of-plane fields. To elucidate the nature of these phases, we develop a minimal anisotropic exchange model. This model, supported by combined first principles calculations and theoretical modeling, quantitatively reproduces our experimental data. In BCAO, the combination of strong bond-independent XXZ anisotropy and geometric frustration leads to significant quantum order by disorder effects that stabilize colinear phases under both zero and finite magnetic fields. 
\end{abstract}

\maketitle

\section{Introduction} \label{sec:intro}

\begin{figure*}
\includegraphics[width=\linewidth]{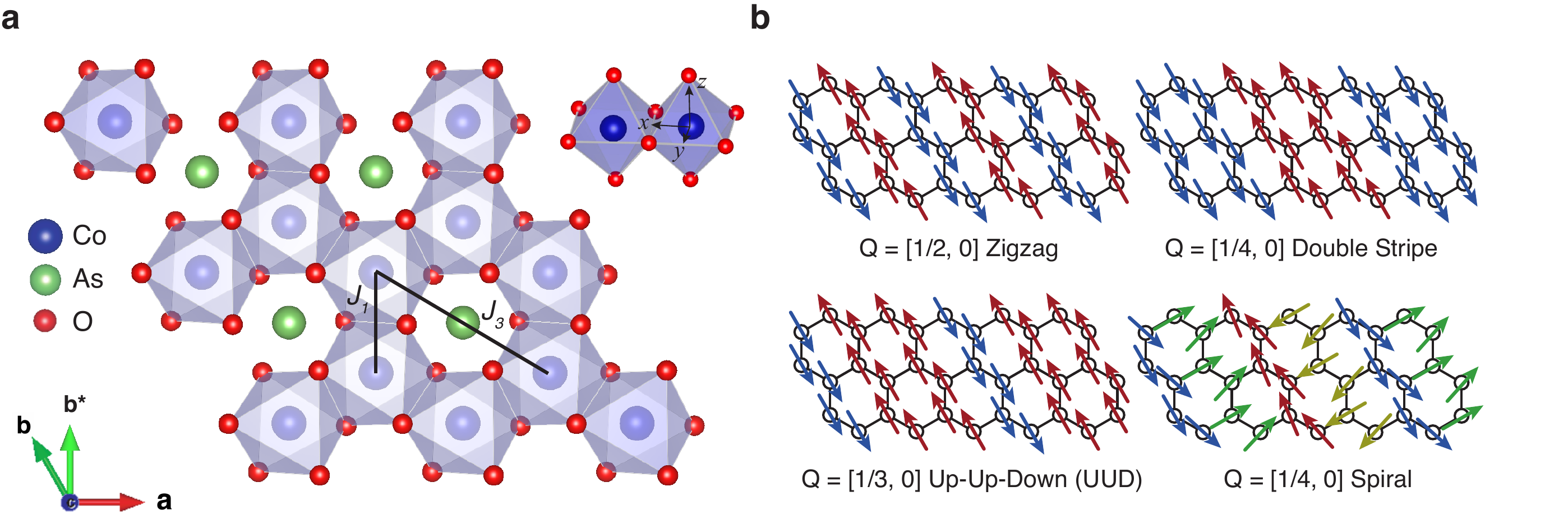}
\caption{\label{fig:schematic} {\bf Crystal and magnetic structures.} {\bf a} Crystal structure of \BCAO{} showing honeycomb magnetic lattice of Co in the $ab$-plane layers. Blue balls denote Cobalt, green balls Arsenic and red balls oxygen. We omitted Ba as they exist between layers. The $a$, $b$, and $c$ are the global coordinate. The $a$ and $b$ point along two zigzag directions of the honeycomb lattice while $b^{*}$ points along the armchair direction. The $c$ points along the out-of-plane direction. The $x, y$ and $z$ are local coordinates used in Eq.~(\ref{Hamil_1}), where the $x$-axis points along the bond-direction while $z$ points along the out-of-plane $c$ axis. The $y$ is perpendicular to the $x$ and the $z$ directions.
$J_{1}$ and $J_{3}$ are the nearest and third nearest interactions, respectively. {\bf b} Energetically competitive magnetic orders at zero and finite field, with in-plane ordering wavevectors ($Q$) indicated.
}
\end{figure*}

The interplay between crystal field symmetry, spin-orbit coupling, and the electron configuration of cobalt ions (Co$^{2+}$, $3d^{7}$) in a nearly ideal octahedral oxygen environment stabilizes the $J_{\rm eff}$~=~1/2 state, a fundamental building block of numerous intriguing physical phenomena in various configurations. One of the intriguing theoretical proposals is that cobaltates with a $J_{\rm eff}$~=~1/2 state can potentially exhibit bond-dependent exchange interactions, including what is lately called the Kitaev exchange interaction \cite{liu2018pseudospin,sano2018kitaev,liu2020kitaev,winter2022magnetic}. It is well known that when the Kitaev exchange interaction dominates the spin Hamiltonian, the ground state can become a Kitaev quantum spin liquid (KQSL) due to strong magnetic frustration \cite{jackeli2008mott,rau2014generic,chaloupka2010kitaev}. The KQSL is highly sought after among frustrated magnets because it has the potential to host exotic emergent phenomena such as Majorana Fermions and topological order, with significant applications in quantum computing \cite{kitaev2006anyons,nayak2008non,kitaev2003fault}. Therefore, several honeycomb cobalt-based magnets, including Na$_{2}$Co$_{2}$TeO$_{6}$ \cite{lin2021field,zhang2023electronic,zhang2024out}, and Na$_{3}$Co$_{2}$SbO$_{6}$  \cite{yan2019magnetic} have been investigated intensively. 

Recently, another cobalt-based honeycomb-lattice material \BCAO{} (BCAO) has been proposed as an ideal platform for realizing a KQSL~\cite{zhong2020phase,zhang2023magnetic} (Figure ~\ref{fig:schematic}a). BCAO shows a series of magnetic phase transitions under applied magnetic field ($H$). The zero-field order is identified as either an incommensurate spiral or double-stripe (DS), which is suppressed in favor of a collinear up-up-down (UUD) phase under field~\cite{Gegenwart2024, regnault1977physica, regnault1979effect, regnault2018polarized}. These spin structures are illustrated in Figure~\ref{fig:schematic}b. For magnetic fields in the crystallographic $ab$-plane, the latter phase is subsequently suppressed by relatively weak fields (0.5~T), above which a KQSL phase has been suggested~\cite{zhong2020phase,tu2022qsl}. For field in the out-of-plane direction, which is theoretically more conducive to a field-induced KQSL phase~\cite{qsl_Hickey2019, qsl_Jiang2019, qsl_Patela2019,gordon2019theory}, a broad magnetic continuum was observed \cite{zhang2023magnetic}. However, conflicting reports exist regarding whether BCAO is best described by an extended Kitaev model with strongly bond-dependent magnetic couplings or as an XXZ-$J_1$-$J_3$ model with weakly bond-dependent couplings; recent inelastic neutron scattering measurements, as well as $ab$ $initio$ calculations~\cite{maksimov2022ab}, align more with the XXZ-$J_1$-$J_3$ model~\cite{halloran2023geometrical}. It has also been pointed out that strong distortions of oxygen octahedra, whose the trigonal term in crystal-field splitting matrix is larger than one fifth of spin-orbit coupling, lead to Ising- and XY-like spin behavior \cite{lines1963magnetic,oguchi1965theory,winter2022magnetic}. Additionally, the weak ligand-assisted exchange interactions weaken bond-dependent exchange interactions \cite{winter2022magnetic}. Hence, further investigation is still required to characterize the phases of BCAO comprehensively. Investigating the angular dependence of phase boundaries is an effective method to understand the origin of magnetic anisotropy \cite{riedl2019sawtooth}. Recently, Safari et al. explored the phase diagram of BCAO using magnetotropic susceptibility measurements and theoretical models~\cite{safari2024quantum}. Their findings indicate that while classical models qualitatively describe the system's behavior, quantum corrections are essential for precise critical field predictions \cite{safari2024quantum}.



In this study, we explored the details of magnetic field-induced ground states in BCAO using ac and dc magnetic susceptibility, capacitance measurement, and torque magnetometry measurements at various temperatures ($T$). With the aim of establishing the origin of the magnetic anisotropy, we study the angular dependent $T$-$H$ phase boundaries by tuning the magnetic field orientation. With the application of a magnetic field, multiple-phase transitions were identified for $H$ along the in-plane direction, consistent with previous studies~\cite{zhong2020phase,zhang2023magnetic,Gegenwart2024,regnault1977physica, regnault1979effect, regnault2018polarized}. In comparison, for $H$ close to the $c$-axis, the critical field for suppression of ordering is drastically increased, and intermediate phase transitions disappear or are washed out at the lowest temperatures.

To understand the full details of the phase diagram, we employ first-principles-based calculations to provide a starting point for analyzing the magnetic couplings. The resulting model is refined by fitting the experimental phase diagram employing second-order energy corrections~\cite{jackeli2015quantum} to account for quantum fluctuation effects on the stability of various magnetic phases. This combined approach allows for the precise reproduction of the experimental observations. Consistent with recent assertions \cite{maksimov2022ab,maksimov2023proximity,jiang2023quantum,safari2024quantum}, we conclude that quantum order-by-disorder (QOBD) effects dominate in BCAO, influencing both the critical field values and specific progression of field-induced phases.

\begin{figure*}
\includegraphics[width=\linewidth]{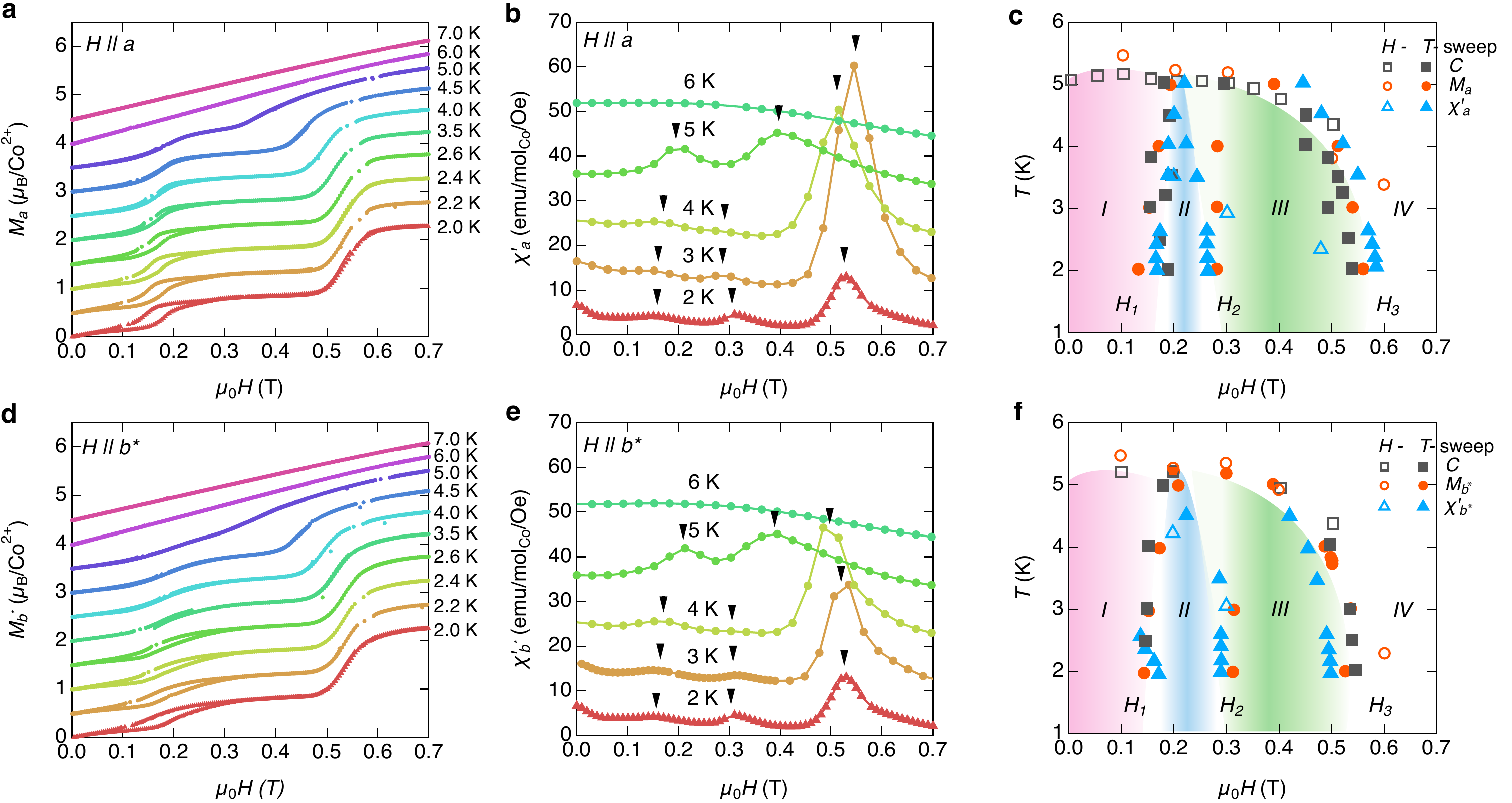}
\caption{\label{fig:maginplane} {\bf Magnetic properties and phase diagrams for in-plane fields.} {\bf a} Magnetization per Co$^{2+}$ at various temperatures with field parallel with $a$. For the clarity, a 0.5~$\mu_{\text{B}}$/Co$^{2+}$ offset was applied between the two curves. {\bf b} Ac susceptibility as a function of field at various temperatures with field parallel with $a$. {\bf c} Magnetic field vs. temperature phase diagrams with field along the $a$-axis. {\bf d} Magnetization per Co$^{2+}$ at various temperatures with field parallel with $b^{*}$. A 0.5~$\mu_{\text{B}}$/Co$^{2+}$ offset was applied between the two curves. {\bf e} Ac susceptibility as a function of field at various temperatures with field parallel with $b^{*}$ direction. {\bf f} Magnetic field vs. temperature phase diagrams with field along the $b^{*}$-axis. Data points are taken from dc magnetization $M$, ac susceptibility $\chi'$, and capacitance measurement $C$ (see Supplementary Figure 2) for the phase diagrams. 
}
\end{figure*}

\section{Results} \label{sec:results}
\subsection{Magnetic properties and phase diagrams for in-plane fields}

Figure~\ref{fig:maginplane}a illustrates the magnetization measured at various temperatures as a function of the field applied along the $a$-axis. Hysteresis loops emerge between 0.1~T ($H_{1}$) and 0.2~T ($H_{2}$), which become smaller with increasing temperature and are almost invisible above 4.0~K. Beyond 0.5~T, there is a steep increase in magnetization before reaching saturation around 0.6~T ($H_{3}$). The saturation magnetization is about 2.3~$\mu_{\text{B}}$/Co$^{2+}$, from which we extracted $g_{ab} \sim 4.6$. The magnetization between the saturation field and $\sim$0.15~T constitutes a one-third magnetization plateau. These findings are consistent with prior studies \cite{regnault1979effect, zhong2020phase, regnault2018polarized}. When the temperature exceeds $T_{\text{N}} \sim$ 5.5~K, all features disappear, and a monotonic increase of the magnetization with field is observed, consistent with a thermal paramagnetic state. Figure~\ref{fig:maginplane}b shows the ac magnetic susceptibility as a function field. The temperature dependent ac magnetic susceptibility is in Supplementary Figure~1. Three peaks are observed: The first two peaks at $\sim$$H_{1}$ and $\sim$$H_{2}$ are consistent with the jump in magnetization in the hysteresis loop and the end of the hysteresis loop. The third peak above $\sim$$H_{3}$ corresponds to the saturation field. ac susceptibility at 6~K does not show any anomalies. The magnetic phase diagram with field along the $a$-axis deduced from these measurements is presented in Figure~\ref{fig:maginplane}c. We observed three thermodynamic phases (I, III, and IV) and one intermediate state (II) due to the hysteresis loop \cite{Gegenwart2024,wang2023single}. Neutron scattering experiments identified phase I as the DS phase or closely related spiral phase with in-plane wavevector close to $Q = [1/4,0]$ \cite{regnault2018polarized,halloran2023geometrical}. Phase III is the UUD phase with 1/3 saturation magnetization and $Q = [1/3,0]$ \cite{regnault1977physica, regnault1979effect, regnault2018polarized}. Given the differing wavevectors and the discontinuous change in wavevector through this phase transition, it is not surprising that these phases are separated by a first order phase transition. Phase II corresponds to a coexistence region of Phase I and III. Lastly, phase IV is the asymptotically polarized paramagnetic phase that is smoothly connected to high temperature paramagnetic phase.

Figures~\ref{fig:maginplane}d-\ref{fig:maginplane}f show analogous results for field along $b^{*}$-axis. For this field direction, we observe the same progression of phase transitions. The hysteresis loops are wider but less pronounced, and $H_3$ is slightly lower in comparison with $H||a$. Therefore, the width of 1/3 plateau (phase III) is slightly narrower with magnetic field along $b^{*}$-axis. 

\subsection{Magnetic properties and phase diagrams for out-of-plane field}

\begin{figure*}
\includegraphics[width=\linewidth]{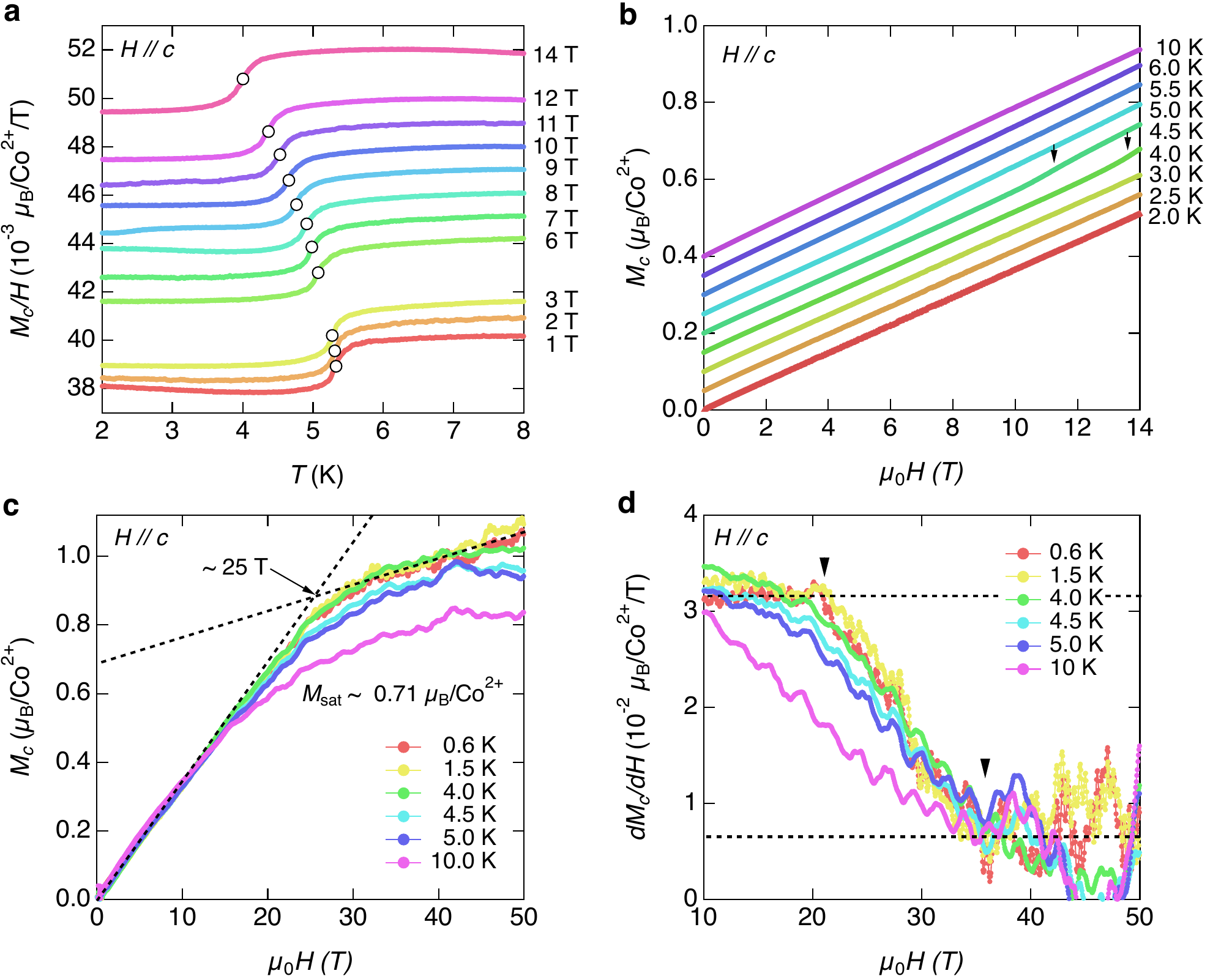}
\caption{\label{fig:magCC} {\bf Magnetic properties for out-of-plane fields.} {\bf a} Temperature dependent DC susceptibility at various magnetic field along out-of-plane direction. An offset of 0.001~$\mu_{B}/$Co$^{2+}$/T multiplied by the field is applied to each curve for clarity. {\bf b} Field dependent magnetization per Co$^{2+}$ measured up to 14 T at various different temperatures, with curves offset by 0.05~$\mu_{B}$/Co$^{2+}$. {\bf c} Pulsed field magnetization up to 50~T at different temperatures. {\bf d} Differentiated magnetization with respect to the field.}
\end{figure*}

Figure~\ref{fig:magCC}a shows the temperature dependent dc magnetic susceptibility ($M_{c}/H$ (T)) measured at various constant external magnetic fields. At 1~T, a clear phase transition from the paramagnetic to DS state is observed and marked with an empty circle around 5.3~K. Raising the field to 14~T only suppresses $T_{\text{N}}$ to 4~K. The phase transition is therefore very robust against out-of-plane magnetic field, which is in sharp contrast with the behavior for in-plane fields. 

Figure~\ref{fig:magCC}b shows the magnetization versus out-of-plane field up to 14~T. At 2~K, the magnetization linearly increases without showing any anomaly, demonstrating a gradual spin canting towards the field direction without phase transitions. At 4~K, we observe a weak slope change around 14~T, denoted as a black arrow, which is consistent with the thermal phase transition between the DS and thermal paramagnet observed in Figure~\ref{fig:magCC}a. With increasing temperature to 4.5~K, the slope change shifts to slightly lower field of about 11.5~T. Suppression of the phase boundary between AFM and paramagnetic phase with increasing temperature is a typical behavior of antiferromagnets. 

Since 14 T is not enough to observe the magnetic saturation, we measured the magnetization with field along c-axis up to 50 T at Pulsed Field Facility in Los Alamos National Laboratory. 
Results from pulsed field magnetization experiments at various temperature are shown in Figure~\ref{fig:magCC}c and ~\ref{fig:magCC}d. The data are admittedly noisy, but the important features are clearly discernible. In the paramagnetic state at 10~K, the magnetization curve resembles the Brillouin function. 
Below $T_{\text{N}}$, the magnetization increases linearly and shows a slope change in the vicinity of 25~T, which we assign as the out-of-plane saturation field. Therefore $dM_{c}$$/dH$ is a constant below $\sim$$20$~T and above $\sim$$35$~T. Due to the small crystal field excitation gap of Co$^{2+}$, it is common to observe a linearly increasing magnetization above the saturation field, due to Van Vleck paramagnetism \cite{lee2014series,susuki2013magnetization}. By extrapolating the Van Vleck contribution to the zero field, we found a saturation moment of about 0.71~$\mu_{B}$/Co$^{2+}$, from which we extract $g_{c}\sim 1.5$. We note that the saturation field with out-of-plane configuration in this study is higher than the field claimed by Zhang et al. \cite{zhang2023magnetic} and Safari et al. \cite{safari2024quantum}. As we demonstrated in Supplementary Figure~4 and 5, a tiny misalignment of the crystal can lead to a drastically suppressed saturation field. For example, only 6$^{\circ}$ degrees off from the $c$-axis suppresses the saturation field 27~T to 4.5~T. 
The 25~T phase transition is smoothly connected to the zero field AFM to paramagnetic phase transition, additionally confirming that it is the saturation field. Thus, the out-of-plane magnetization measurements demonstrate that BCAO is highly anisotropic between in-plane and out-of-plane fields, with a saturation field ratio $H_{sat}^{c}/H_{sat}^{a}$ of about 50.

\subsection{Angular dependence of magnetic phase transitions}
\begin{figure*}
\includegraphics[width=\linewidth]{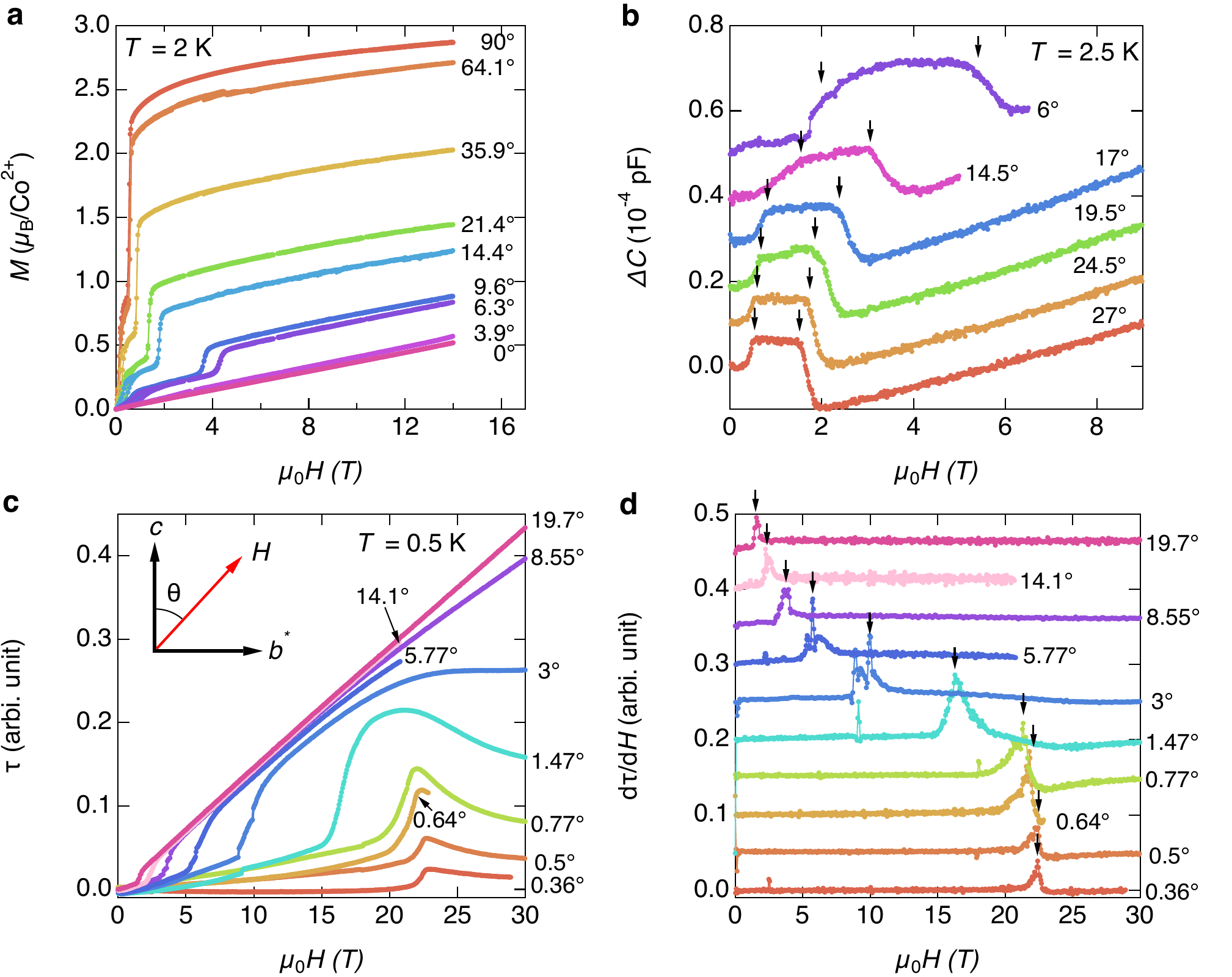}
\caption{\label{fig:magangle} {\bf Field-angle dependence of the magnetic properties.} {\bf a} Magnetization per Co$^{2+}$ as a function of field at various angles at $T$ = 2~K. 0$^{\circ}$ is $c$-axis and 90$^{\circ}$ is $b^{*}$-axis. {\bf b} Magnetic field dependent capacitance with electric field along $b^{*}$-axis at different angles at $T$ = 2.5~K. For clarity, we applied the offset of $1\times 10^{-5}$~pF between two curves. The arrows indicate the phase transition fields. {\bf c} Field dependent magnetic torque at various angles at $T$ = 0.5~K. {\bf d} Differentiated magnetic torque with respect to the magnetic field at various angles with an offset. The arrows indicate the peak positions. All these measurements were performed with sweep-up magnetic field.}
\end{figure*}
To investigate how the phase boundaries evolve with the field direction, we measured the magnetization, capacitance, and magnetic torque below $T_{\text{N}}$ while varying the magnetic field orientation across the $c$-$b^*$ plane. In Figure~\ref{fig:magangle}a, we first show the evolution of the magnetization. For an out-of-plane field ($\theta = 0^\circ$), the magnetization does not show any features up to 14~T. However, with a small rotation of the field direction towards the $b^{*}$ direction, the transitions to the intermediate UUD and polarized phases appear. At $\theta = 6.3^{\circ}$, the phase transitions to the UUD and polarized phases occur at 1~T and 4~T, respectively. These move towards $\sim$$H_{1}$ and $\sim$$H_{2}$ as the field direction approaches $b^{*}$-axis, consistent with Figure~\ref{fig:maginplane}d. The magnetic moment also monotonically increases as the field is rotated towards b$^{*}$ direction, consistent with the anisotropy of the $g$-values.

\begin{figure*}[t]
\includegraphics[width=\linewidth]{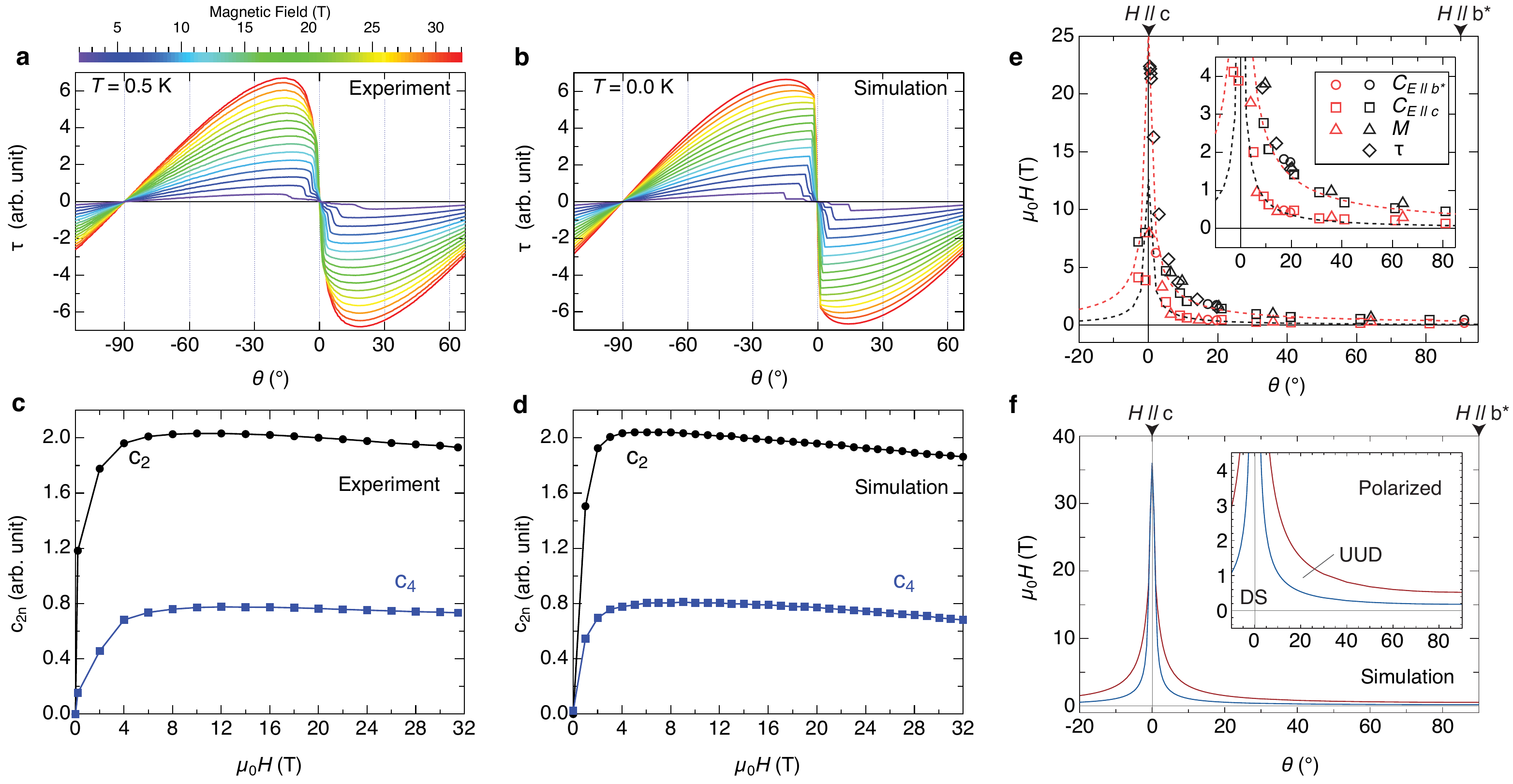}
\caption{\label{fig:torque} {\bf Experimental and theoretical angle-dependence magnetic torque and phase diagrams.} {\bf a} $b^*c$-plane magnetic torque as a function of angle for various magnetic fields measured up to 32~T at $T = 0.5$ K. {\bf b} Theoretical simulation of torque on the basis of second-order corrected state energies employing optimal model discussed in the text. {\bf c} Coefficients $\left|c_{2}\right|$ and $\left|c_{4}\right|$ as a function of magnetic field extracted from 0.5~K experimental data. {\bf c} The same coefficients extracted from 0.0~K theoretical simulations. {\bf e} Critical magnetic fields as a function of field angle in the $b^*c$-plane. Data points were obtained through the analysis of the magnetization, capacitance, and torque measurements. The red and black symbols correspond to the $H_{1}$ and $H_3$, respectively. The dashed lines are guides for the eye. {\bf f} Theoretical phase diagram for optimized model (see text).}
\end{figure*}

Capacitance and magnetostriction measurements can be sensitive probes of magnetic phase transitions as the change in spin structure accompanies the change in electric susceptibility and lattice parameter of magnetic insulators. We found this to be the case for \BCAO{}. Figure~\ref{fig:magangle}b shows the change in capacitance as a function of magnetic field at various angles. We applied the electric field at 16 kHz along $b^{*}$-axis. At 27$^{\circ}$, the capacitance change shows clear two phase transitions corresponding to the phase transition from DS to UUD, and UUD to polarized phase. They evolve to higher fields as the field becomes close to $c$-axis. The anomaly from the phase transition from DS to UUD phase becomes weaker while that from UUD to polarized phase is more or less similar at all angles. The capacitance change and magnetostriction data with in-plane magnetic field can be found in Supplementary Figure~2 and 3.

Figure~\ref{fig:magangle}c and~\ref{fig:magangle}d show magnetic torque ($\tau$) and its differentiation with magnetic field ($d\tau/dH$) up to 30 T as a function of magnetic field at different fixed angles. Both $\tau$ and $d\tau/dH$ display a clear anomaly at phase transitions. When the field is only at 0.36$^{\circ}$ from $c$-axis, a sharp increase in $\tau$ and a peak $d\tau/dH$ appear at about 23~T. This is reasonably well consistent with the saturation field observed in pulsed field magnetization shown in Figure~\ref{fig:magCC}c. The anomaly shifts quickly to lower field as the angle becomes larger than 1$^{\circ}$ and ends up being only 2~T at 19.7$^{\circ}$. 

In Figure~\ref{fig:torque}a, we show the magnetic torque measured for continuous field-angle sweeps between the $c$ and $b^{*}$ axes at $T$ = 0.5~K at various fixed magnitudes of $H$. All curves exhibit 180$^{\circ}$ periodicity and a saw-tooth shape due to the strong magnetic anisotropy. The sharp change of sign of the torque near $0^{\circ}$ indicates the $c$-axis is the hard axis. Additional step-like anomalies are observed near $\theta = 0^\circ$ marking the first order transitions $H_1$ and $H_3$. These two features are only clearly observable up to 10 T and they merge into one above 10~T. 

In general, the magnetic torque may be expanded as \cite{riedl2019sawtooth}:
\begin{equation}
    \frac{\tau}{H} = \sum_{n = 1}^{\infty} c_{2n}\sin{(2n\theta)} = c_{2} \sin(2\theta) + c_{4}\sin(4\theta) + \cdots
\end{equation}
The experimental field-dependence of $\left|c_{2}\right|$ and $\left|c_{4}\right|$ is plotted in Figure~\ref{fig:torque}c. Both coefficients increase steeply up to 4~T and then slowly decrease above 4~T up to the maximum field. The ratio of $c_{2}$ to $c_{4}$ remains more or less constant at $\sim$$0.4$ across the entire field range due to the dominant saw-tooth curve shapes.  In the limit of large field, such that the majority of the field angles correspond to the polarized phase, the $c_{2n}$ coefficients have two contributions. The first contribution is field-independent, and depends exclusively on the $g$-anisotropy. For the implied range of $g$-values ($g_{ab} \sim 4.6$, $g_{c} = 1.5 - 2.5$), the limiting ratio for this contribution leads to $|c_4/c_2| = 0.17 - 0.27$. The next order contribution to the $c_{2n}$ scales as $1/H$, and is related to both the XXZ exchange anisotropy and \mbox{$g$-anisotropy} \cite{riedl2019sawtooth}.  The findings that $c_2$ and $c_4$ decay with increasing field, and that their ratio exceeds the $g$-only value both demonstrate significant exchange anisotropy is necessary to explain the observed torque. Below, we develop an exchange model for BCAO, and use these observations to verify the model.

From the phase transitions observed in the angular dependence of magnetic properties in Figure~\ref{fig:magangle} and \ref{fig:torque}a, an angular dependence of the magnetic phase diagram is constructed, as shown in Figure~\ref{fig:torque}e. The red empty symbols represent the phase transition from the DS to the UUD phase ($H_{1}$), and the black empty symbols indicate the saturation field ($H_{3}$). As the field direction moves from the $b^{*}$-axis to the $c$-axis, both phase boundaries sharply increase below 10$^{\circ}$. The saturation field reaches above 23~T, whereas the phase boundary between the DS and the UUD merges with the saturation field within $5^{\circ}$ above about 10~T.

\subsection{Microscopic magnetic model}
To first define a realistic range of magnetic couplings, we employ the des Cloizeaux effective Hamiltonian (dCEH) approach \cite{des1960extension,riedl2019ab,dhakal2024spin}, based on exact diagonalization of the electronic $d$-orbital Hamiltonian for clusters include one and two Co sites [see Methods]. 

We first estimate the $g$-tensor by diagonalizing the electronic Hamiltonian on one Co site, and projecting the Zeeman operator into the low-energy space of $j_{1/2}$ states. This yields $g_{ab} = 5.1$ and $g_{c} = 2.4$, which further confirms accurate reproduction of the trigonal crystal field splitting in BCAO. To estimate the magnetic couplings, we exactly diagonalize the electronic Hamiltonian on pairs of sites and project the resulting low-energy Hamiltonian onto $j_{1/2}$ states. The relatively low symmetry crystalline environment allows for a large number of independent exchange constants. In particular, in the $R\bar{3}$ space group, each first and third neighbor bond is constrained only by inversion symmetry, allowing six independent exchange parameters per bond:

\begin{align}
\label{Hamil_1}
    \mathcal{H} =& \  \sum_{ij} J_{ij}^{xy} (S_i^x S_j^x + S_i^y S_j^y) + J_{ij}^{z} S_i^zS_j^z  \nonumber \\ & \ + 2J_{ij}^{\pm\pm}(S_i^xS_j^x - S_i^y S_j^y) + \Gamma_{ij}^{yz} (S_i^y S_j^z + S_i^z S_j^y) \nonumber \\ & \ + \Gamma_{ij}^{xz}(S_i^x S_j^z + S_i^z S_j^x) + \Gamma_{ij}^{xy}(S_i^y S_j^x + S_i^x S_j^y)
\end{align}
where the bond-dependent local $x$-axis is oriented parallel to the bond, and the $z$-axis is parallel to the crystallographic $c$-axis. Second neighbor couplings tend to be weaker and are ignored in first approximation. By varying the strengths of the Coulomb interactions within a realistic range [see Methods], we establish reasonable ranges for the first and third neighbor couplings: 
$J_1^{xy} \sim -8$ to $-12$ meV, 
$J_1^z/J_1^{xy} \sim 0.5$ to 0.6, 
$J_1^{\pm\pm} \sim -0.6$ to $-0.8$ meV, 
$\Gamma_1^{yz} \sim 0.35$ to 0.5 meV,
$J_3^{xy}/J_1^{xy} \sim -0.25$ to $-0.4$, and 
$J_3^z/J_3^{xy} \sim 0.2$. The remaining couplings are all found to have magnitude smaller than 0.1 meV.

There are several aspects of the computed couplings that bear mentioning. In BCAO, the local trigonal distortion of the CoO$_6$ octahedra has a significant impact on the magnetic Hamiltonian, inducing both the large $g$-anisotropy, and bond-independent XXZ anisotropy of the intersite exchange. Microscopically, the third neighbor couplings are dominated by a single exchange path involving the $e_g$ electrons, which leads to antiferromagnetic interactions that are rendered anisotropic only by the crystal field effects. As such, the off-diagonal anisotropies should be nearly vanishing and the third neighbor interactions should be parameterized by a pure XXZ anisotropy, with $J_3^{xy}$ being the same sign as $J_3^z$. For the first neighbor couplings, the dominant anisotropy is also of the bond-independent XXZ type. While Co$^{2+}$ materials can, in principle, host strongly bond-dependent couplings \cite{sano2018kitaev,liu2018pseudospin,liu2020kitaev,winter2022magnetic}, in BCAO we find that these are suppressed by a combination of factors. First, the large trigonal distortion from a pure octahedral environment significantly modifies the composition of the $J_{\rm eff}$ = 1/2 moments, and imposes significant XXZ anisotropy on all couplings instead of bond-dependent exchange interactions. Second, the largest contribution to the first neighbor exchange comes from ferromagnetic ``charge-transfer'' contributions arising from hybridization of the Co $e_g$ orbitals with the oxygen $p$-orbitals. Unlike the couplings between $t_{2g}$ electrons, these interactions do not display large bond-dependent anisotropies. As such, the bond-dependent anisotropies ($J_1^{\pm\pm}$ and $\Gamma_1^{yz}$) are subleading. Although some of these features are found in previously proposed models for BCAO \cite{halloran2023geometrical,maksimov2022ab,maksimov2023proximity,safari2024quantum}, not all are reproduced. This raises the question of whether the phase diagram can be understood by employing a model that follows more closely these microscopic considerations. 

\begin{figure}[t]
\includegraphics[width=\linewidth]{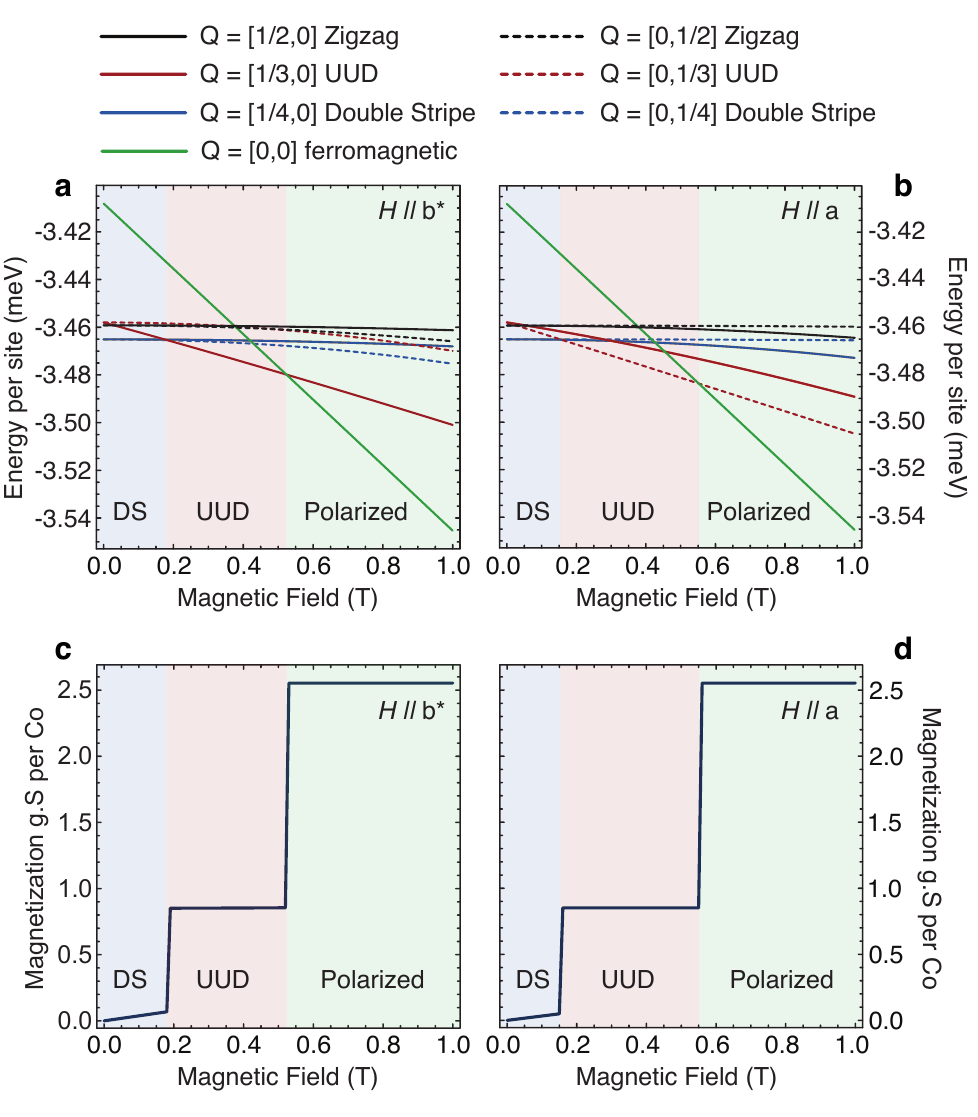}
\caption{{\bf Energy per site calculations for the energetically competing phases.} {\bf a}, {\bf b} Comparison of 2nd order corrected energies per site for different magnetic states for fields $H||a$ and $H||b^*$, respectively. {\bf c},{\bf d} Theoretical magnetization per site for different field directions showing 1/3 magnetization plateau.}
\label{fig:optimal_energies}
\end{figure}

To address this question, we made a comprehensive search of the parameter space around the predicted range, with a focus on fitting the experimental phase diagram. BCAO necessarily lies in a parameter region with multiple competing phases, making it necessary to consider quantum corrections to the classical state energies. We employ the method outlined in Ref.~\onlinecite{jackeli2015quantum} [see Methods], which considers states defined by classical spin vectors, but accounts for local fluctuation effects on the state energies at second order in perturbation theory. This naturally includes the leading quantum order-by-disorder (QOBD) effects. From this search, we propose a simplified, but extremely finely tuned, six-parameter model:
$J_1^{xy} = -11.75$, $J_1^{z} = -5.89$, $J_1^{\pm\pm} = -0.33$, $\Gamma_1^{yz} = -0.15$, $J_3^{xy} = +3.37$, and $J_3^{z} = +0.79$ meV. The parameters are more consistent with microscopic considerations but otherwise broadly compatible with previously proposed models for BCAO with dominant XXZ anisotropy \cite{maksimov2022ab,safari2024quantum}.

In Figure~\ref{fig:optimal_energies}, we show the computed evolution of the state energies and magnetization per site of the above model for in-plane fields. We evaluate $m = |\sum_i g\cdot\mathbf{S}_i|$ from the classical spin vectors for which the second order energies correspond. The zero-field ground state is found to be a colinear DS phase, with spins oriented in the plane, as shown in Figure~\ref{fig:schematic}b. Under applied field, there are two first order transitions. The first is to the magnetization plateau UUD state, and the second to the polarized state. We estimate for $H \ || \ a$: $H_{1} = 0.16$ T, $H_{3}= 0.55$ T, and for $H \ || \ b^*$: $H_{1} = 0.18$ T, $H_{3} = 0.52$ T. Thus, the model reproduces the slight anisotropy in the in-field critical fields observed in experiment. The predicted magnetization as a function of field is shown in Figure~\ref{fig:optimal_energies}c and \ref{fig:optimal_energies}d, and reproduces well the experimental magnetization curves in Figure~\ref{fig:maginplane}.

In Figure~\ref{fig:torque}e and ~\ref{fig:torque}f, we show a comparison of the experimental and theoretical phase diagram for fields oriented in the $b^*c$-plane. For strictly out-of-plane fields $H||c$, the model reproduces the direct transition from the DS state to the polarized phase. The critical field depends very strongly on the field angle; for $\theta = 0^\circ$, we compute $H_3 = 36$~T, while for $\theta = 1^\circ$, we compute $H_3 = 18$~T. This range is comparable to the broadened saturation observed in the pulsed-field measurements in Figure~\ref{fig:magCC}.
The sharp increase of the critical fields in the vicinity of the $c$-axis is mainly due to the large bond-independent XXZ anisotropy, which leads to an easy $ab$-plane. Overall, the model reproduces the phase boundaries well, confirming that the specific progression of field-induced phases can be understood within a microscopically compatible model. 

Finally, to further check that the model reproduces the characteristic magnetic anisotropy of BCAO, we simulated the magnetic torque data from the second-order corrected energies, where $\tau = \partial E/ \partial\theta$. Results are shown in Figure~\ref{fig:torque}b alongside the experimental data. The simulations reproduce very well both the positions and magnitudes of the first order jumps in $\tau$ corresponding to the various phase boundaries, as well as the rounding of $\tau(\theta)$ at high field. A precise comparison can be made on the basis of the coefficients $c_{2n}$. In Figure~\ref{fig:torque}d, we show the theoretical evolution of these coefficients with field. The finding that $c_2$ and $c_4$ are well reproduced by the model implies the relative magnitudes of both the $g$-anisotropy and XXZ exchange anisotropy are well captured. 

Overall, we conclude that the microscopically motivated model provides a very good reproduction of the experimental phase diagram of BCAO and therefore may serve as a starting point for future refinement. The present results confirm the validity of recently proposed models with primarily bond-independent XXZ anisotropy \cite{maksimov2022ab,halloran2023geometrical,safari2024quantum}, but show that the complex phase diagram of BCAO illuminated by various recent studies is compatible with the microscopic expectations for this material. 

\section{Discussion}
As a consequence of the exchange anisotropy and frustration between first and third neighbor couplings, the phase diagram of BCAO is uniquely dominated by QOBD effects. Over the region of couplings we considered for BCAO, we observe several trends when comparing classical state energies with the second order corrected energies. First, primarily due to the strong XXZ anisotropy, both antiferromagnetic and polarized (ferromagnetic) states have finite energy corrections. Similar to the DMRG studies in Ref. \cite{jiang2023quantum,safari2024quantum}, for the physically relevant magnitudes of the XXZ anisotropy, we find that fluctuations tend to stabilize the polarized phase over the other phases. 
Overall, we find that quantum fluctuations tend to reduce the critical fields by favoring the polarized state, consistent with the conclusions of \cite{safari2024quantum}. Second, we emphasize that the collinear DS and UUD phases only appear as ground states due to QOBD effects. We did not find any realistic parameters for which these are classical ground states for any magnetic field. Conversely, the spiral phase is never found to be stable once fluctuation effects are included. These findings are consistent with \cite{maksimov2023proximity,jiang2023quantum}. Finally, we emphasize that the present model for BCAO is extremely finely tuned. As can be anticipated from the (necessarily) small differences of the state energies in Figure~\ref{fig:optimal_energies}a and ~\ref{fig:optimal_energies}b deviations of any coupling on the order of $\pm 0.01$ meV may be sufficient to ruin the agreement with experiment. This sensitivity likely also leads to variations in the results of different numerical approaches. In this work, we opted to use the inexpensive approximation of second-order energy corrections to facilitate a fine parameter search. Future refinement of the model could employ more sophisticated numerical methods, provided those approaches adequately treat the QOBD effects.

In conclusion, this work is a comprehensive study of the magnetic field-induced ground states of BCAO, employing various techniques such as dc~magnetization, capacitance, $ac$ magnetic susceptibility at low temperatures. We mapped out the full $T$-$H$ phase diagram for different magnetic field orientations, including $H$ along $a$, $b^*$, $c$, and $c'$. To understand the specific progression of field-induced phases, we employed first-principles-based calculations to derive a model that reproduces all essential features of the low-temperature phase diagram. The stabilization of two colinear phases (I and III) in both zero and finite fields is attributed to the significant influence of QOBD effects. BCAO presents an intriguing material where quantum fluctuation effects dominate the phase diagram at both zero and finite field. This investigation highlights new avenues for understanding the complex behavior of BCAO under external magnetic fields.

\section{Methods} \label{sec:methods}
\subsection{Experimental details}

High-quality single crystals of BaCo$_2$(AsO$_4$)$_2$ synthesized through a flux method at Oak Ridge National Laboratory in Oak Ridge, for which the detailed methods are described elsewhere ~\cite{halloran2023geometrical}. At low temperature, dc~magnetization ($M$) is measured up to 60~T through millisecond timescales in pulsed magnet using custom-built $^3$He pumping (400~mK–3~K), and ac magnetic susceptibility ($\chi_{ac}$) and capacitance ($C$) measurements are shown here performed in PPMS for 2~K–30~K under the magnetic fields up to 14~T at Pulsed Field Facility in Los Alamos. For dc~magnetization and ac magnetic susceptibility measurements, the sample with 1~mm$^2$ was aligned along $a$, $b^*$, $c$, and $c'$ and mounted on the sample holder inside the compensation and pick-up coil, where $c'$-axis is 6$^{\circ}$ from $c$-axis on the $b^*$-$c$ plane.
The electrical contacts were made by painting silver epoxy on the two parallel opposite surfaces for the capacitance measurement and Andeen-Hagerling AH-2700A capacitance bridge was employed to measure the capacitance.

Magnetic torque was measured with the 31 T resistive magnet at the National High Magnetic Field Laboratory in Tallahassee, Florida using a home-made torque magnetometer. The torque magnetometer was formed from two parallel BeCu plates. A tiny aligned crystal with the mass of 0.13 mg was attached using epoxy on the cantilever plates. The capacitance change was monitored while the field or angle was swept using an AH-2700A capacitance bridge. More details can be found in https://nationalmaglab.org/user-facilities/dc-field/measurement-techniques/torque-magnetometry-dc/. 

We then converted the measured capacitance to the deflected angle ($\alpha$) using the following equation
\begin{equation}
    C(\alpha) = \frac{\epsilon_{0}\epsilon_{r}w}{\tan \alpha} \ln {\frac{l\tan\alpha-h_{0}}{h_{0}}},
\end{equation}
where $\alpha$ is the defected angle that is proportional to the magnetic torque, $\epsilon_{0}$ and $\epsilon_{r}$ are vacuum and relative permittivity, respectively, and $w$, $l$, and $h_{0}$ are width, length of the BeCu plate and the gap distance between two BeCu plates at zero magnetic field. We extracted the coefficients $c_{2n}$ using the following equation
\begin{equation}
    c_{2n} = \frac{2}{\pi} \int_{\theta_{1}}^{\theta_{i}+\pi} \frac{\tau}{H} \sin(2n\theta) d\theta,
\end{equation}
where $\theta_{i}$ is the initial angle, $\tau$ is the measured magnetic torque signal.

\subsection{Theoretical phase diagrams}

In order to account for quantum fluctuation effects, we employ the method of second-order energy corrections outlined in Ref.~\onlinecite{jackeli2015quantum}. At each site $i$, the local quantization axis (denoted $\hat{z}_i^\prime$) is along the local spin vector, such that $\langle S_i^{z\prime}\rangle = +\frac{1}{2}$ for all sites. Fluctuations are then driven by terms acting like $J_{ij}^{--}S_i^{-\prime}S_j^{-\prime}$. The second-order energy correction due to such fluctuations is:

\begin{align}
\Delta E^{(2)} = \sum_{{\rm bonds } \  ij} \frac{J_{ij}^{++} J_{ij}^{--}|\langle m| S_i^{-\prime}S_j^{-\prime}|gs\rangle|^2}{E_{gs,{\rm classical}} - E_{m, {\rm classical}}}
\end{align}
where $E_{gs,{\rm classical}}$ is the classical energy of the spin configuration, and $E_{m,{\rm classical}}$ is the classical energy of the state with the two spins along the bond $ij$ flipped. For the example of a Heisenberg honeycomb antiferromagnet, the classical ground state energy is $E_{gs,\rm classical} = -0.375 \ J$ per site, whereas the present approach provides $E_{gs,\rm classical} + \Delta E^{(2)} = -0.563 \ J$ per site. The latter estimate (which includes local singlet fluctuations) is in much better agreement with $E_{gs} = -0.544 \ J$ per site from series expansion \cite{oitmaa1992quantum} and quantum Monte Carlo \cite{low2009properties}. 
In order to capture the energetically competitive magnetic orders depicted in Figure~\ref{fig:schematic}b, we employ a combination of 4-, 6-, and 8-site clusters of spins with appropriate boundary conditions, and numerically minimize $E_{\rm classical} + \Delta E^{(2)}$ with respect to the spin orientations. 
We consider the commensurate spiral state with $Q = [1/4, 0]$ as a proxy for the proposed incommensurate spiral with $Q = [0.27,0]$ \cite{halloran2023geometrical}. We find the $Q = [1/4, 0]$ spiral to be the classical ground state for a range of parameters, but it is completely replaced by the colinear ferromagnetic and $Q = [1/4, 0]$ DS phases once $\Delta E^{(2)}$ is included. 

\subsection{First principles calculations}


To estimate the magnetic couplings, we consider an electronic model for the Co $3d$-orbitals, which is a sum of, respectively, one- and two-particle terms: $\mathcal{H}_{\rm el} = \mathcal{H}_{1p}+\mathcal{H}_{2p}$. The one-particle terms include intersite hopping, intrasite crystal field, and spin-orbit coupling, $\mathcal{H}_{1p} = \mathcal{H}_{hop}+\mathcal{H}_{\rm CF} + \mathcal{H}_{\rm SO}$:

\begin{align}
    \mathcal{H}_{hop} =& \  \sum_{ij\alpha\beta\sigma}t_{ij}^{\alpha\beta} c_{i,\alpha,\sigma}^\dagger c_{j,\beta,\sigma}
    \\
    \mathcal{H}_{\rm CF} = & \ \sum_{i\alpha\beta\sigma}d_{i}^{\alpha\beta}c_{i,\alpha,\sigma}^\dagger c_{i,\beta,\sigma}
    \\
    \mathcal{H}_{\rm SO} = & \ \sum_{i\alpha\beta\sigma\sigma^\prime} \lambda_{\rm Mn} \langle \phi_i^\alpha(\sigma)|\mathbf{L}\cdot\mathbf{S}|\phi_{i}^\beta(\sigma^\prime)\rangle c_{i,\alpha,\sigma}^\dagger c_{i,\beta,\sigma^\prime}
\end{align}
where $c_{i,\alpha,\sigma}^\dagger$ creates an electron at Co site $i$, in $d$-orbital $\alpha$, with spin $\sigma$. To parameterize the one-particle terms, we performed fully relativistic density functional theory (DFT) calculations at the GGA (PBE \cite{perdew1996generalized}) level using FPLO \cite{opahle1999full}, a 12$\times$12$\times$12 $k$-grid, and the structure from Ref.~\onlinecite{djordevic2008baco2}. The resulting Kohn-Sham Bloch functions were projected onto Co $d$-orbitals to yield single-particle terms from the Kohn-Sham Fock matrix written in the Wannier basis \cite{koepernik2023symmetry}. For BCAO, this approach yields accurate crystal-field excitation energies \cite{mou2024comparative}, confirming the validity of the computed single-particle parameters. 

For the two-particle terms, we consider a combination of on-site and nearest neighbor intersite contributions: $\mathcal{H}_{2p} = \mathcal{H}_U + \mathcal{H}_{J}^{\rm nn}$, respectively. The on-site contributions are given by:

\begin{align}
\mathcal{H}_U = \sum_{i\alpha\beta\delta\gamma}\sum_{\sigma\sigma^\prime}U_{\alpha\beta\gamma\delta} \ c_{i,\alpha,\sigma}^\dagger c_{i,\beta,\sigma^\prime}^\dagger c_{i,\gamma,\sigma^\prime} c_{i,\delta,\sigma}
\end{align}

In the spherically symmetric approximation \cite{sugano1970multiplets}, $U_{\alpha\beta\gamma\delta}$ are parameterized by the Slater parameters $F_0^{dd}, F_2^{dd}, F_4^{dd}$. For the (screened) on-site metal Coulomb interactions, we use $F_2^{dd} = 11.55$ eV, and $F_4^{dd} = 7.68$ eV, following the fitted parameters for LiCoO$_2$ from Ref.~\onlinecite{van1991electronic}. We consider $U_{t2g}= F_0^{dd} + (4/49)(F_2^{dd}+F_4^{dd})$ within the realistic range of 5.0 to 5.5 eV.

The nearest neighbor intersite Hund's coupling is given by:

\begin{align}
    \mathcal{H}_{J}^{\rm nn} = \sum_{ij\alpha\beta\sigma\sigma^\prime} J_{H,ij}^{\alpha\beta} \ c_{i,\alpha,\sigma}^\dagger c_{j,\beta,\sigma^\prime}^\dagger c_{i,\alpha,\sigma^\prime}c_{j,\beta,\sigma}
\end{align}

A significant contribution to the intersite magnetic couplings comes from ligand ``charge transfer'' exchange terms \cite{liu2018pseudospin,winter2022magnetic}, which account for perturbative processes where two holes meet on a single oxygen ligand. When downfolded into the $d$-orbital Wannier basis, these appear as intersite Hund's coupling terms. They give an overall ferromagnetic contribution to the intersite magnetic couplings between the low-energy $j_{1/2}$ doublets, which may be rendered anisotropic (bond-independent XXZ anisotropy) by the crystal-field splitting of the $t_{2g}$ orbitals. In a previous study of BCAO \cite{maksimov2022ab}, some of the authors accounted for this term via an estimated correction added to the computed couplings. Here, we employ the improved approach utilized in \cite{dhakal2024hybrid,dhakal2024spin,konieczna2024understanding}  that captures these effects at the level of the electronic Hamiltonian. The $d$-orbital Wannier functions can be written:

\begin{align}
    |\tilde{d}_{i,\alpha,\sigma}\rangle = \phi_{i,\alpha}^0|d_{i,\alpha,\sigma}\rangle + \sum_{n,\beta} \phi_{i,\alpha}^{n,a}|p_{n,a,\sigma}\rangle
\end{align}

where $i$ labels a given Co metal site, $\alpha$ is a $d$-orbital, $n$ labels an oxygen ligand site, and $a\in\{x,y,z\}$ is a $p$-orbital. The tilde on the left side of the equation indicates a Wannier function, whereas the absence of a tilde indicates a bare atomic orbital. Rotating the $p$-orbital Coulomb terms into the $\tilde{d}$ Wannier basis results in residual intersite Hund's coupling given by:

\begin{align}\label{eqn:intersiteHund}
    J_{H,ij}^{\alpha\beta} =  \sum_n \sum_{\alpha\beta}  & \ 
(U^p-J_H^p)\ \phi_{i,\alpha}^{n,a}\phi_{i,\alpha}^{n,b}\phi_{j,\beta}^{n,a} \phi_{j,\beta}^{n,b}  \nonumber
\\ & \ 
+ J_H^p \ \phi_{i,\alpha}^{n,a}\phi_{i,\alpha}^{n,a} \phi_{j,\beta}^{n,b}\phi_{j,\beta}^{n,b}  
\end{align}

where $U^p$ and $J_H^p$ are the atomic ligand Hubbard and Hund's couplings, respectively. The computed magnetic couplings depend most strongly on the overall magnitude of $J_{ij}^{\alpha\beta}$, rather than the specific orbital dependence, which provides for some inconsequential flexibility in the choice of $U^p$ and $J_H^p$.  
We take $U^p = 2J_H^p$, and 
consider $J_H^p$ within the range \blue{of} 0.4 to 0.6 eV. This choice corresponds to roughly 60\% to 80\% of the atomic value for oxygen \cite{anno1971systematic}. The sum over oxygen sites in Eq.~ (\ref{eqn:intersiteHund}) is then evaluated from the {\it ab-initio} Wannier function coefficients for the bridging oxygen atoms for each bond. 

\section*{Acknowledgements} \label{sec:acknowledgements}
The authors would like to thank A. L. Chernyshev, P. A. Maksimov, Y. B. Kim, F. Desrochers, E. Z. Zhang, K. A. Modic, and S. A. Crooker for insightful discussions. 
The experimental work, analysis and manuscript writing  (S.L., S.Z., S.M.T., L.P., C.A.B., V.S.Z., and M.L.) was funded by the US Department of Energy, Office of Science, National Quantum Information Science Research Centers. The theoretical work, analysis and manuscript writing of S. M. W. was funded by the Oak Ridge Associated Universities
(ORAU) through the Ralph E. Powe Junior Faculty Enhancement Award. Computations were performed using the Wake Forest University (WFU) High-Performance Computing
Facility \cite{WakeHPC}, a centrally managed computational resource available to WFU researchers including
faculty, staff, students, and collaborators. A portion of this work was performed at the National High Magnetic Field Laboratory, which is supported by National Science Foundation Cooperative Agreement No. DMR-2128556, the State of Florida, and the U.S. Department of Energy.

\section*{Author contributions} \label{sec:Authorcontributions}
V.S.Z. and M.L. conceived and designed the project. S.L., S.Z., S.M.T., E.S.C. and M.L. performed measurements. S.M.W. performed the theoretical calculations.
L.P. and C.A.B. prepared high quality samples. S.L., V.S.Z., S.M.W., and M.L. wrote the manuscript, with contributions from all authors.

\section*{Competing interests} 
The authors declare no competing interests.\\

{\bf Correspondence} and requests for materials should be addressed to
Vivien S. Zapf, Stephen M. Winter or Minseong Lee.
\setcounter{figure}{0}
\renewcommand{\figurename}{Supplemental Figure}
\clearpage

\section{Supplementary material \--- \\ ``Quantum Order by Disorder: A Key to Understanding the Magnetic Phases of \BCAO{}''}

\begin{figure}[h]
\includegraphics[width=1\columnwidth]{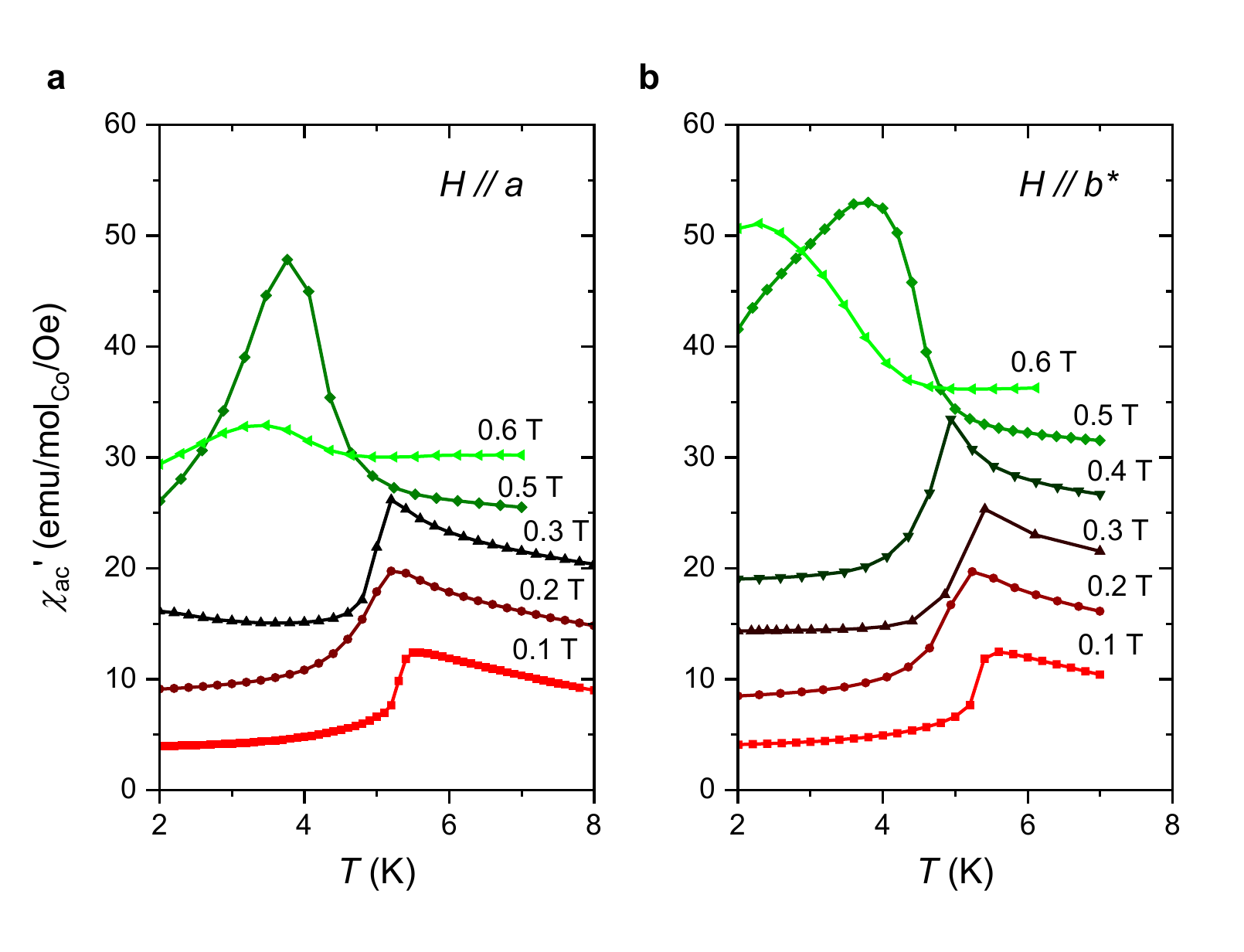}
\caption{\label{fig:XacTforHab} {\bf In-plane magnetic susceptibility of \BCAO{}}  Temperature-dependent ac magnetic susceptibility at various magnetic field along in-plane direction with offsets for magnetic field along $a$ ({\bf a}) and $b^{*}$-axis ({\bf b}), respectively.}
\end{figure}

\begin{figure}[h]
\includegraphics[width=1\columnwidth]{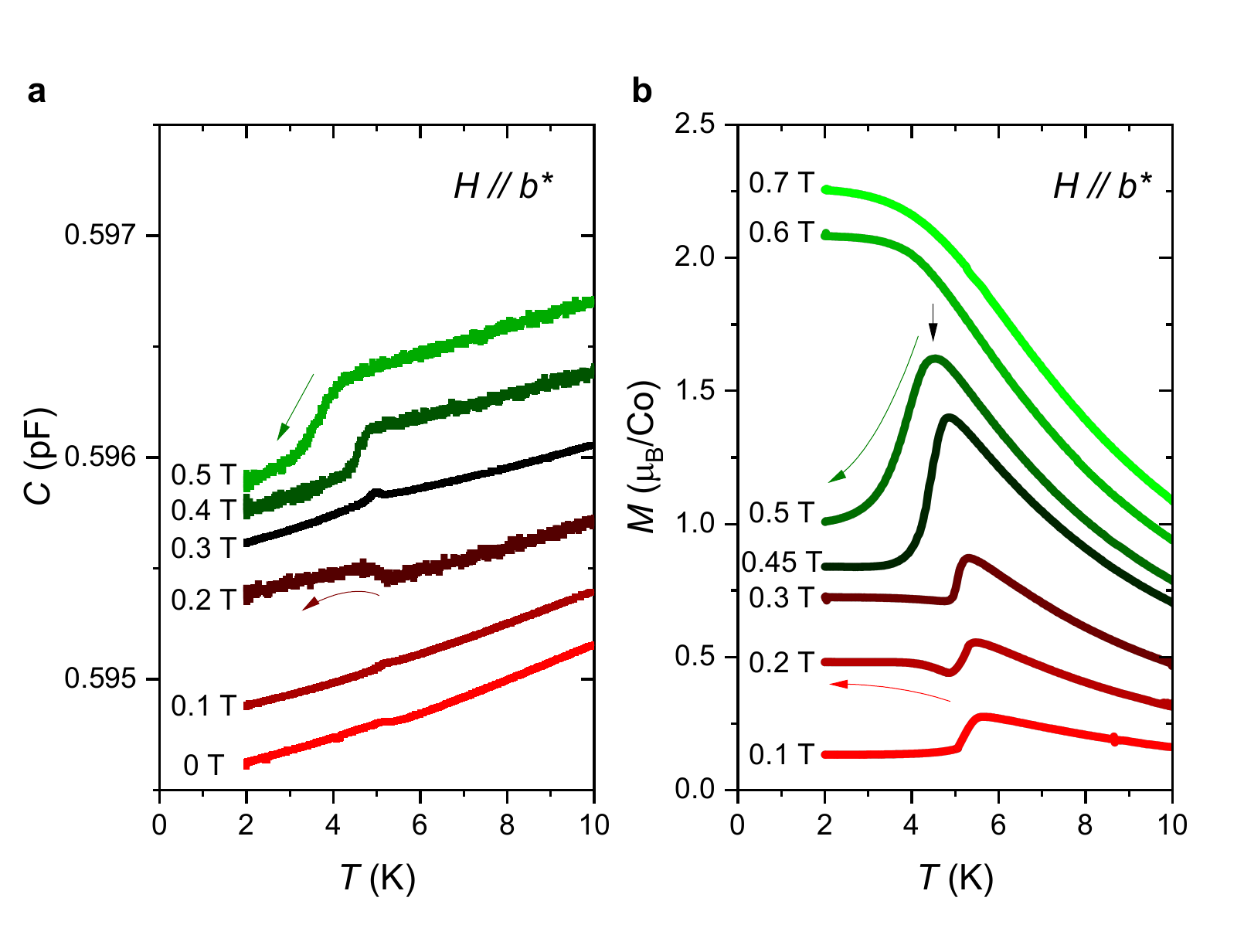}
\caption{\label{fig:CTandMTTforHb} {\bf In-plane capacitance and magnetization of \BCAO{}} Temperature-dependent capacitance ({\bf a}) and magnetization ({\bf b}) at various magnetic field with offsets for magnetic field along $b^*$-axis. The ac electrical field excitation is applied along $c$-axis.}
\end{figure}

\begin{figure}[h]
\includegraphics[width=1\columnwidth]{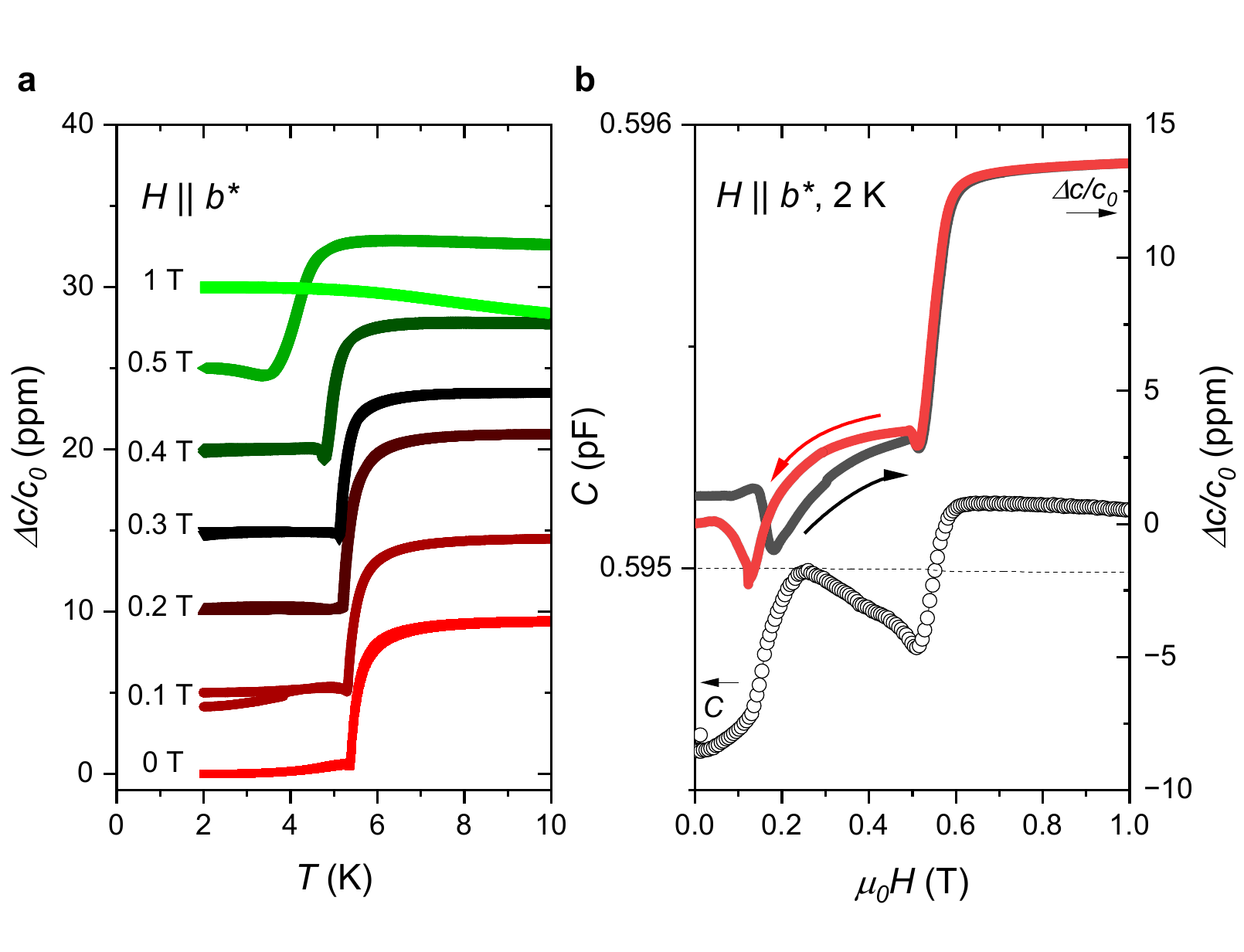}
\caption{\label{fig:deltaccTandCHforHb} {\bf Thermal expansion and magnetostriction for $H||b^*$} {\bf (a)} Normalized thermal expansion along $c$-axis at various magnetic field for $H||b^*$. {\bf (b)} Magnetic field-dependent capacitance (open circles) and normalized magnetostriction along $c$-axis (solid lines) are plotted together at 2~K for $H||b^*$. For capacitance, electrical field is applied along $c$-axis.}
\end{figure}

\begin{figure}[h]
\includegraphics[width=1\columnwidth]{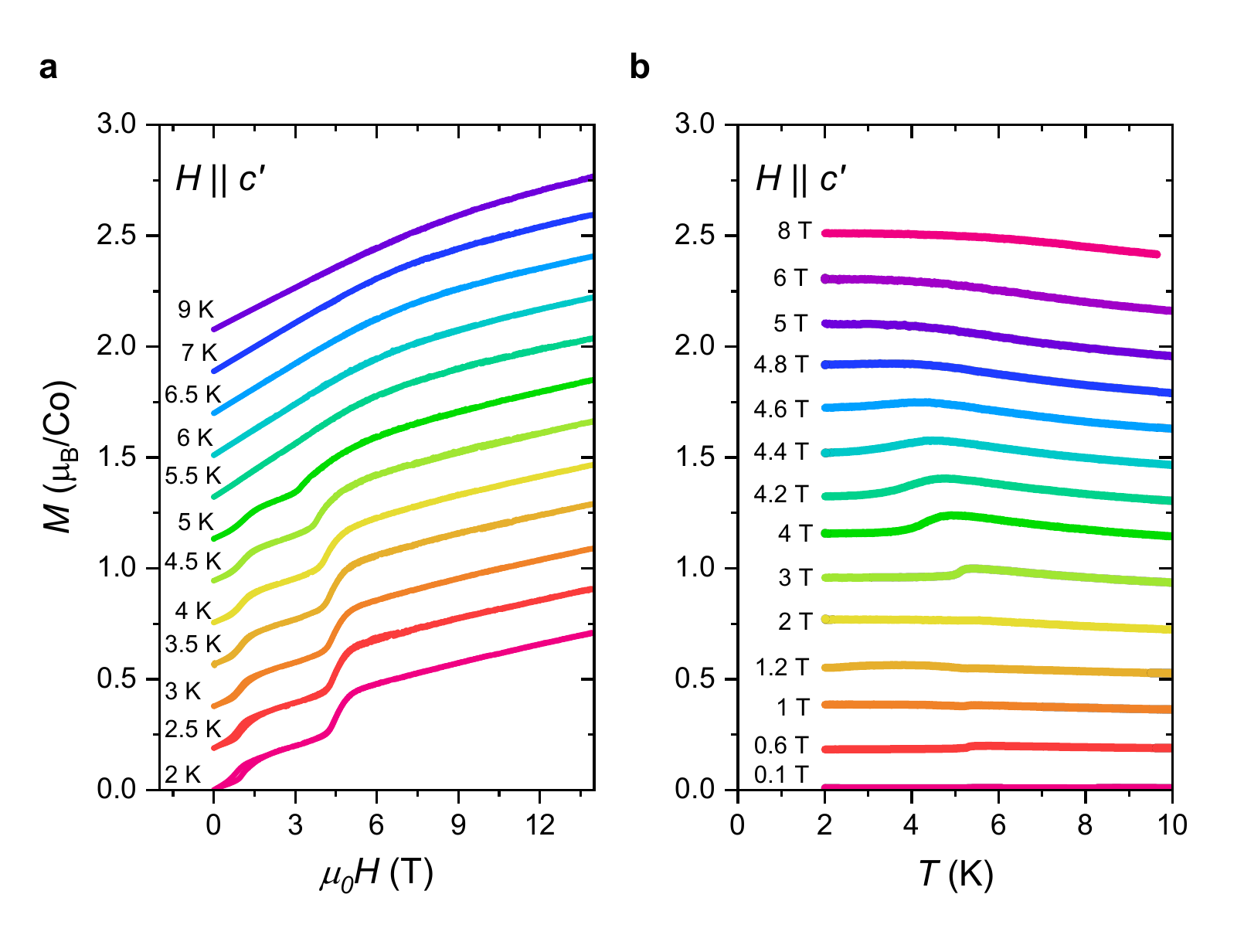}
\caption{\label{fig:MTandMHforHcprime} {\bf Magnetization per Co$^{2+}$ for H$||c'$, 6$^\circ$ away from $c$-axis on $b^*c$~plane.}  {\bf (a)} Magnetization at various temperatures with field {\bf (b)} Magnetization with field along $H||c'$ at various magnetic field with temperature.}
\end{figure}


\begin{figure}[h]
\includegraphics[width=1\columnwidth]{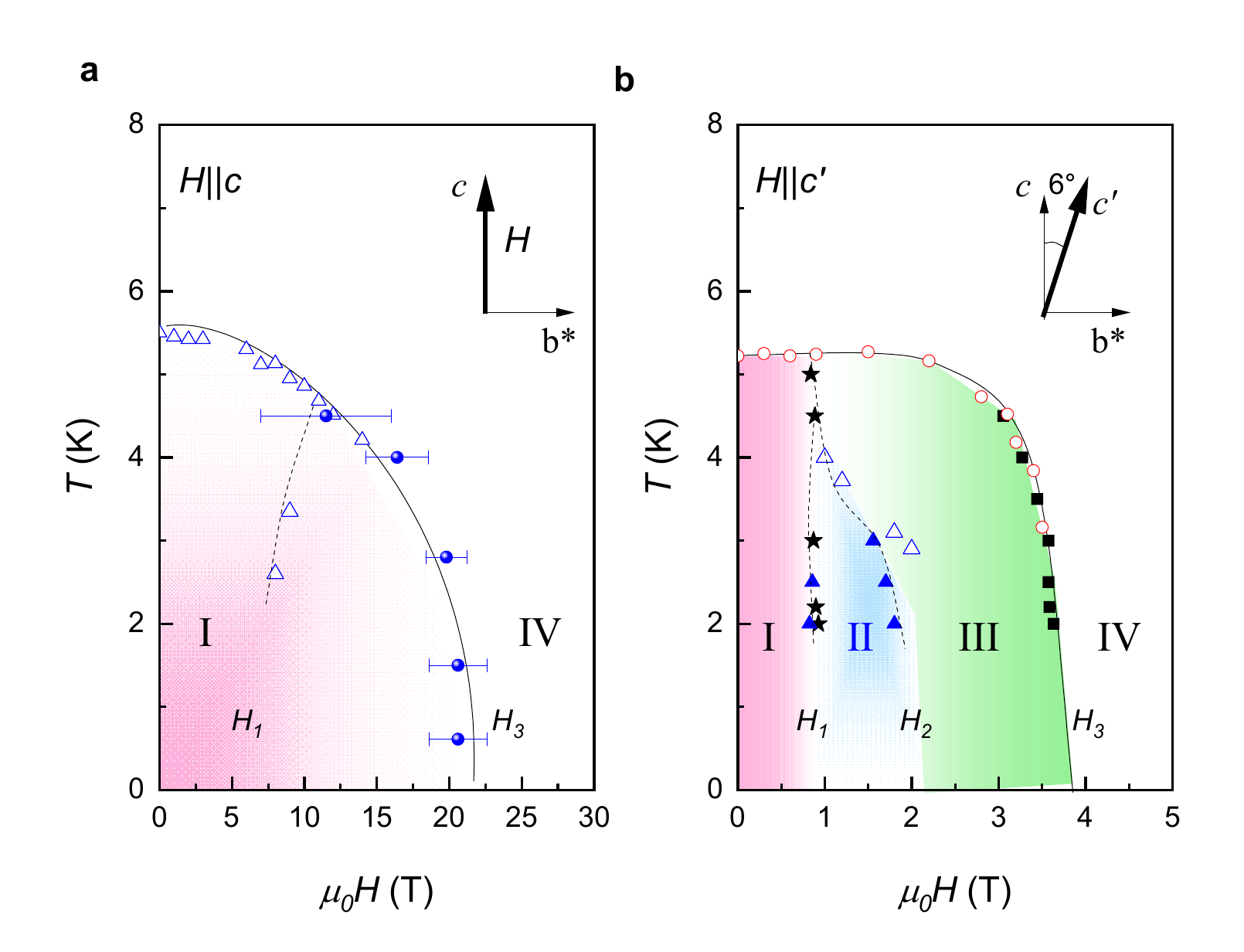}
\caption{\label{fig:phasediagramCandCprime} {\bf Phase diagram for H$||$ $c$ and $c'$. $c'$-axis is 6$^\circ$ away from $c$-axis on $b^*-c$~plane.} Magnetic field vs. temperature phase diagrams with field along $c$-axis {\bf (a)} and $c'$-axis {\bf (b)}, respectively. Data points are taken from dc magnetization $M$ (blue symbols), ac susceptibility $\chi$ (red symbols), and capacitance measurement $C$ (black symbols).}
\end{figure}
\clearpage

\bibliography{main}

\end{document}